\documentclass{article}
\usepackage[T1]{fontenc}
\usepackage{xcolor}
\usepackage{amssymb}
\usepackage{verbatim}
\newcommand{\hidden}[1]{{\color{gray}#1}} % version 1

\usepackage{tikz-qtree}
\usepackage{pst-tree}

\tikzset{every tree node/.style={minimum width=2em,draw,circle},
         blank/.style={draw=none},
         edge from parent/.style=
         {draw, edge from parent path={(\tikzparentnode) -- (\tikzchildnode)}},
         level distance=1.5cm}

\usepackage{setspace}
\usepackage{tikz}
\usepackage{pgfplots}

\title{The Hidden Binary Search Tree:\\
{\large A Balanced Rotation-Free Search Tree in the AVL RAM Model}}
\author{Saulo Queiroz \\ Email: {sauloqueiroz@utfpr.edu.br} \\
\\ Academic Department of Informatics \\ 
Federal University of Technology (UTFPR)\\ Ponta Grossa, Brazil}

\begin{document}
\maketitle
\begin{abstract}
In this paper we generalize the definition of ``Search Trees'' (ST) to enable
reference values other than the key of prior inserted nodes. The idea builds
on the assumption an $n$-node AVL (or Red-Black) requires to assure $O(\log_2n)$ worst-case
search time, namely, a single comparison between two keys takes constant time. This means 
the size of each key in bits is fixed to $B=c\log_2 n$ ($c\geq1$) once $n$ is determined, otherwise
the $O(1)$-time comparison assumption does not hold. Based on this we calculate
\emph{ideal} reference values from the mid-point of the interval $0..2^B$. This idea follows 
`recursively' to assure each node along the search path is provided a reference value that
guarantees an overall logarithmic time. Because the search tree property works only when keys
are compared to reference values and these values are calculated only during searches,
we term the data structure as the Hidden Binary Search Tree (HBST). We show elementary functions  
to maintain the HSBT height $O(B)=O(\log_2n)$. This result requires no special order on the input 
-- as does BST -- nor self-balancing procedures, as do AVL and Red-Black.
\end{abstract}

\section{Introduction}
%A Search Tree data structure is a set of keys recursively defined in 
%such a way that the key of any node
A Search Tree (ST) evolves upon insertions/deletions according to
the nodes' key field value. Such field works as a kind of \emph{reference}
to help placing a new unique value $k$ in the tree rooted by the key value $k_r$. 
For a Binary Search Tree (BST), $k$ is placed to the left or right subtrees 
of $k_r$ if $k<k_r$ and $k>k_r$, respectively.
\emph{This definition forces the reference value to always be the key field of 
previously inserted nodes}. Hence, if a sequence consisted of $n$ `bad' 
keys is given as input, the resulting 
worst-case search performance is linear, missing the opportunity
to build a $O(\log_2 n)$-height tree. 

A classical way to preventing
BST to unbalance consists in walking the insertion/deletion path back to the 
root to check/update height-related informations and trigger node rotation(s) 
if necessary, the so-called self-balanced BSTs e.g. AVL~\cite{avl-62}, ``Red-Black''~\cite{redblack-78}.
Those tasks increase programming complexity in comparison to BSTs. Also, 
they might cause the self-balanced BSTs to perform worse than a common BST 
when the sequence of insertion keys leads the latter to be ``naturally'' balanced.
Indeed, Knuth~\cite{artprogramming-vol3-knuth-1998} shows 
that BSTs requires only $2\ln n\approx 1.386\log_2n$ comparisons if keys are inserted 
in a random order.% An AVL insertion in such tree would be about twice worse than in 
%a BST because BST does not need to adjust balancing factors after finding node's position.
For this dilemma, he  suggests a `balanced attitude' considering
self-balanced trees for large $n$ -- due to the BST's `annoying (linear) possibility' --
and BSTs for lower $n$ because of its reduced overhead and simpler programming. 
%He completes saying that ``well-balanced trees are common, and degenerate trees are very rare''.

In this paper we wonder about the condition(s) under which (if any) it is possible to design  
a rotation-free BST with $O(\log_2 n)$-height regardless of the $n$-size sequence of insertion 
keys. To achieve that we generalize the definition of ``Search Tree'' by enabling \emph{reference} 
values other than the key values of prior inserted nodes. We refer to them as \emph{hidden reference 
values} because \emph{they guide search procedures but requires no kind of permanent registration}.
To derive the proposed's tree worst-case 
search, insertion and deletion time complexities we consider the Random-Access Machine (RAM)
model~\cite{cook-rammodel-1972} under the same assumptions taken by  AVL and Red-Black.  With that,
after the maximum number $n$ of keys in the tree is determined, the maximum number of bits $B$ to represent 
a key is computed as a constant bounded to $\lceil\log_2(n)\rceil$. Otherwise, if $n$ is not known in advance,
 $B$ can not be assumed
as constant. Thus, the constant time assumed for comparison between keys ``clearly becomes an 
unrealistic scenario''~\cite{cormen-2009} and the $O(\log_2 n)$ complexity no more hold for the
self-balanced BSTs.

%Under this assumption, the number of bits
%to uniquely represent any key in a \emph{given} range $[0,n-1]$ can be bounded to $B=\lceil\log_2(n)\rceil$. 
%If, differently, $B$ can grow arbitrarily, a simple comparison between two keys take a $\omega(1)$-time 
%and the $O(\log_2n)$ search time of self-balanced BSTs no more hold because the variable $B$ must account.
Given an input key, we rely on $B$ to calculate hidden reference values for each node 
over the search path. These values correspond to the \emph{ideal} sequence of insertion keys
to build a balanced BST with at most $n$ nodes. Because new incoming keys are placed 
in the tree based on those reference values -- rather than on the values of prior inserted keys-- 
the ``search tree'' property does not hold. However, the resulting tree can be viewed as a ``search 
tree'' in the sense that the 
\emph{hidden reference value} of an arbitrary node is always greater (less) than any \emph{key value} 
in its own left (right) subtree\footnote{The insertion case where the input key is equal to the hidden
reference value is a matter of design choice.}. For this reason we term this 
data structure as the Hidden Binary Search Tree (HBST). We present elementary algorithms to maintain the
 HBST's height bounded to  $O(B)=O(\log_2 n)$. The algorithms assume neither special order keys nor any 
kind self-balancing rotation procedure.

The reminder of this paper is organized as follows. 
In  Section \ref{sec:base} we discuss the basic idea behind the HBST.
In Section \ref{sec:codigo} and \ref{sec:analise} we present the elementary
functions and their worst-case complexities, respectively. Finally in Section \ref{sec:conclusao}
we summarize this work and discuss future directions.
%worst-case height is $\leq \lfloor \log_2(2^B) +1\rfloor$.
%Since there is no need to update any kind of ``balancing factors'', HBST can couple `the better 
%of the two worlds': if the insertion sequence of $n$ keys is `bad' (e.g. increasing), HBST's height 
%`degenerates' up to $B+1$ e.g., in a universe of  $10^{100}$ (1 googol) keys, height is bounded $334$ 
%levels while BST's is $10^{100}$. On the other hand, if the insertion sequence is `naturally good', 
%HBST walk trough the search path once since it does not need to update any kind of balancing factor.

%{AVL says \color{red} a very large number of information is not required to scan or record new info...}

% PARA N PEQUENO, É LINEAR. PORÉM

%figure i
\section{The Reference Hidden BST Algorithm}\label{sec:base}
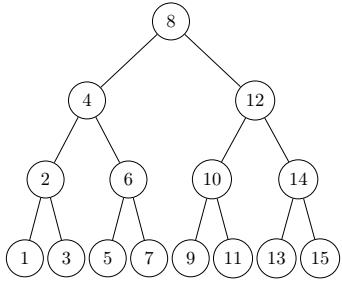
\begin{figure}[h!]
\centering
\caption{BST recursively built from mid-point intervals. Starting point $[1,15]$.}
\label{fig:base}
\begin{tikzpicture}[scale=0.7]
\centering
\Tree
[.8
    [.4
      [.2
         [.1 ]
         [.3 ]         
      ]
      [.6
         [.5 ]          
         [.7 ]         
       ]
    ]
    [.12
      [.10
         [.9 ]
         [.11 ]         
      ]
      [.14
         [.13 ]          
         [.15 ]         
       ]     
    ]
]
\end{tikzpicture}
\end{figure}

Let $n$ be the number of nodes of a BST, each of which uniquely identified by a key $k$
from the integer interval $[0,n-1]$\footnote{$k\in[0,n-1]=\{k\in \mathbb{N}|0\leq k\leq n-1\}$.}. 
In several practical scenarios, $n$ is determined in advance when the data structure programmer determines
 a data type with $B$ bits for the nodes' key field. From the interval $[0,2^B]$, one can build 
a balanced BST following an idea reminiscent to the Merge sort algorithm. This resulting
BST is illustrated in Fig.~\ref{fig:base} for $B=4$. Firstly, 
the algorithm takes the interval $[0,2^B]$ as input and choose its mid-point
$k_m=\lfloor(0+2^B)/2\rfloor$ to insert in the BST. The same idea applies recursively to the root's 
left and right subtrees with the intervals $[a,k_m]$ and $[k_m, b]$, respectively. Note that
$a=0$ and $b=2^B$ in the first iteration.

The idea just described may not seem to be valuable because the insertion 
sequence is not known \emph{a priori}. However, one can benefit from it if the 
``search tree'' property can be relaxed (actually generalized) to include 
reference values other than prior inserted keys. These reference values need 
not to be stored in nodes. They can be computed, for example, taking as reference an ideal 
insertion sequence to guide which subtree the search must follow in 
each iteration (or recursive call). Then, for a given reference search value, 
all nodes at its left subtree have value 
less than it  whereas all nodes in the right subtree have
values greater than it. The insertion case where the input key is equal to the hidden
reference value is a matter of design choice. A binary tree that satisfy the search 
property this way we name as Hidden Binary Search Tree (HBST).

\begin{figure}[h!]
\centering
\caption{HBST built from the insertion sequence $0,1,2,3,4,\dots,15$. $B=4$. 
The upper and lower bounds interval values are passed from one iteration to another; 
their \underline{mid-point} is the hidden search reference value. The interval signs are merely illustrative.}
\label{fig:hbst}
\begin{tikzpicture}[scale=0.7]
\centering
\Tree
[.\node [label=right:\hidden{[0,\underline{8},16[}] {0};
    [.\node [label=right:\hidden{]0,\underline{4},8[}] {1};
      [.\node [label=right:\hidden{]0,\underline{2},4[}] {2};
        \edge[blank];\node[blank]{}; 
         [.\node[label=right:\hidden{]2,\underline{3},4[}] {3}; ]
      ]
      [.\node [label=right:\hidden{[4,\underline{6},8[}] {4};
         [. \node [label=right:\hidden{]4,\underline{5},6[}] {5};]          
         [.\node [label=right:\hidden{[6,\underline{7},8[}] {6}; 
           \edge[blank];\node[blank]{}; 
            [.\node [label=right:\hidden{[7,\underline{7},8[}] {7};]
         ]         
       ]
    ]
    [.\node [label=right:\hidden{[8,\underline{12},16[}] {8};
      [.\node [label=right:\hidden{]8,\underline{10},12[}] {9};
                    \edge[blank];\node[blank]{}; 
         [.\node [label=right:\hidden{[10,\underline{11},12[}] {10}; 
           \edge[blank];\node[blank]{}; 
         [.\node [label=right:\hidden{[11,\underline{11},12[}] {11}; ]           
         ]         
      ]
      [.\node [label=right:\hidden{[12,\underline{14},16[}] {12};
         [. \node [label=right:\hidden{]12,\underline{13},14]}] {13};]          
         [.  \node [label=right:\hidden{[14,\underline{15},16[}] {14};
            \edge[blank];\node[blank]{}; 
         [.\node [label=right:\hidden{[15,\underline{15},16[}] {15}; ]           
         ]          
       ]     
    ]
]
\end{tikzpicture}
\end{figure}
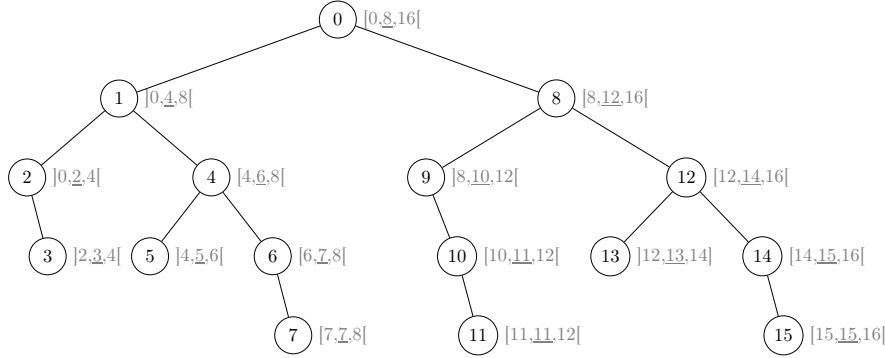

An HBST built from the insertion sequence $0,1,2,\dots,15$ with $B=4$ is illustrated  
in Fig.~\ref{fig:hbst} where values equal to the reference value are insert to the right.
The first insertion is the trivial case. After that, the second 
input consists of the key $k=1$ along with the 
interval\footnote{$k\in[0,n[ =\{k\in \mathbb{N}|0\leq k< n\}$.} $[0,2^{4}[$. 
The algorithm check that there is a node in the current level (the root, in this case), 
and calculates the hidden search reference value $k_m$ (shown in the center of the node's 
interval) from the given interval doing
$\lfloor (0+16)/2\rfloor$. The same idea applies recursively to the root's left and right 
subtrees with the intervals $[a,k_m[$ and  $[k_m, 2^B[$, respectively.
Note that $a=0$ in the first iteration and the interval signs are merely illustrative.

\subsection{HBST's Search Property}
In BST or variants thereof, the ST property is always checked considering the \emph{same
field} of different nodes, usually the key field. That said, it is clear the ST property does
not hold in HBST, as one can easily see in Fig.~\ref{fig:hbst}. However, the hidden reference 
tree associated to the reference values of the interval $0\dots 2^B$, does satisfy the property. 
Besides that, and most important, if $h_{ref}$ is found to be the root's hidden reference value in the HBST 
(sub)tree $T_H$, then \emph{the HBST's insertion rule for an arbitrary key $k$  mandates that 
$k$ must be inserted to the left subtree of $T_H$ if $k<h_{ref}$ or to the right, otherwise}.

%Specially recommended for insertion/deletion intensive applications.
\section{HBST: First Elementary Functions in C}\label{sec:codigo}
In this Section we present the first elementary functions \texttt{insert},
\texttt{search} and \texttt{lazyDel} for inserting, searching and deleting
a given input key in the HBST. The deletion function employs a lazy strategy: 
the node is only removed logically (key field assigned to flag \texttt{-1}) such that 
the space can be reused later by the insertion function. Without loss of generality,
the insertion function assumes the given new key is not already in the tree.

A  `hard deletion' function (not shown here) works just like in standard BSP unless 
the node to be removed has two children. In HBST there is no need to find the minimum 
from right subtree nor maximum from left subtree: the substitute can be any descendant 
leaf node. We choose mnemonic name for the nodes' fields just as 
in a typical BST. The remainder set of assumptions for them are embedded as comment in 
the code itself. 

All functions calculate the hidden reference values considering the quantity of bits
implied by the data type chosen for the key field, a $32$-bit integer in the case.
Since the interval to calculate the hidden reference value halves from one recursive 
call (or iteration) to another, the size of the interval decreases at least by one order of
magnitude e.g., $2^{32}$, $2^{31}$, $\cdots$, $2^0$. This assures the number of iterations
is bounded by the number of bits of the chosen data type. \emph{One variation of the insertion
algorithm may consider calculating a specific upper-interval per iteration instead of
passing them across iterations} (recursions). In this case, if the root subtree in the
$i$-th iteration is $k_i$ then its hidden upper-bound is the largest value possible to 
generate with the minimum number of bits required to represent $k_i$, i.e. $2^{\lceil \log_2(k_i)\rceil}$.
With this a new node can be inserted in-between prior inserted nodes.

\begin{verbatim}
/* Assumptions:
 *  unique key values, HBSTNode typical BST structure, r is 
 *  valid ref pointer to the root, alocateNewNode is
 *  a function to allocate and connect new node.
 *  keys are signed 32-bit integer (negatives discarded) 
 *  i.e. B = 32, n<=2^32. First call: min=0, max=2^32, in C: 1<<32. 
 */
#include <stdlib.h>
HBSTNode *insert(HBSTNode **r, int newKey, unsigned int min, unsigned int max)
{
  if (*r==NULL)
    //alocate new node as in a BST. Return pointer to it or NULL
    return alocateNewNode(r, newKey); 

  // OPTIONAL: make use of space released by function lazyDel
  // for simplicity we assume newKey is not currently in the tree.
  if ((*r)->key == -1)
   { (*r)->key == newKey; return *r; }

  unsigned int hiddenRef = (min + max)/2;
  if (newkey < hiddenRef)
    return insert(&((*r)->left), newKey, min, hiddenRef);
  else
    return insert(&((*r)->right), newKey, hiddenRef, max);
}


HBSTNode *search(HBSTNode *r, int key, unsigned int min, unsigned int max)
{
  if (r == NULL || min > max) return NULL;
  if (r->key == key)
    return r; 
  unsigned int hiddenRef = (min + max)/2;
  if (key < hiddenRef)
    return search(r->left, key, min,hiddenRef);
  else
    return search(r->right, key, hiddenRef, max);
}

/*
 * Assume dynamic environment. Avoid expensive
 * memory management. Employ mem reuse.
*/
HBSTNode *lazyDel(HBSTNode *r, int key, unsigned int min, unsigned int max)
{
  HBSTNode *killMe = search(r, key, min, max);
  if (killMe == NULL || r->key == -1) return NULL;
  killMe->key = -1;
  return killMe;
}
\end{verbatim}

\section{Complexity}\label{sec:analise}
The worst-case order of growth $T(n)$ for the HBST performance
is dominated by the search strategy common to all functions presented
in Section~\ref{sec:codigo}.
 $T(n)$ can be readily obtained by the recurrence equation (\ref{eq:basic}), 
where $n$ is the input parameter \texttt{max}. Recall that we are assuming the RAM 
model~\cite{cook-rammodel-1972} in which the word size can not grow arbitrarily after 
$n$ is chosen~\cite{cormen-2009}. This is the same assumption under which AVL and Red-Black 
get running time is $O(1)\times O(\log_2n)$, where the first term represent the constant time
to perform a comparison between integers and the second the height of tree.
As in binary search, AVL and Red-Black, each round of HBST function solves approximately 
half the input size at a $O(1)$ time cost. Once the number of bits $B$ is assigned to the
key field, maximum asymptotic height of HBST is logarithmic on the input (Eq.~\ref{eq:resultN}) or,
alternatively, linear on $B$ (Eq.~\ref{eq:resultB}). This performance requires no kind of balancing procedure 
nor ``good'' insertion sequences.

%     \begin{itemize}
      \begin{eqnarray} \label{eq:basic}
        T(n) &=&
%        \left\{
%          \begin{array}{lcl}
%                 2  \mbox{, se } n \leq 1\\
             	 c + T(n/2) \mbox{, if } n>1 \\
 %         \end{array}
 %       \right.\\
           T(n) &=& O(\log_2 n)  \label{eq:resultN} \\
           T(n) &=& O(\log_2 2^B) \nonumber\\
           T(n) &=& O(B) \label{eq:resultB}
      \end{eqnarray}

\subsection{Practical Considerations}

The hidden tree underlying an HBST is nothing but a balanced tree composed of all values from
$0..n$, where $n=2^B-1$. Since the resulting tree is balanced, its height is $O(\log_2 n)$. This
reveals the three worst-case is bounded to $O(\log_2 2^B)= O(B)$. Considering a practical example 
in which the key field is declared as a 64-bit integer, the data structure supports no more than
$2^{64}-1\approx 10^{19}$ distinct keys and a comparison between two keys takes $O(64)=O(1)$.
For any quantity $n'<2^{64}$, a balanced BST has its complexity bounded to $O(\log_2 n')$ 
while HBST is bounded to $O(B)$, i.e. $\approx 65$ iterations to find/insert/delete a key in this case. 
Thus, HBST's performance may degrade if, for example, $B$ grows on demand i.e. the value of the 
(really) largest key can not be estimated in advance. In this case, a single comparison takes 
$\omega(1)$ time and some kind of technology to increase $B$ on-demand is required.

\section{Conclusion and Future Work}\label{sec:conclusao}
In this work we showed that it is possible to relaxe the definition of
search tree while keeping almost unchanged the main elementary functions 
of a typical Binary Search Tree (BST). We achieved that by generalizing
the ``search tree'' property allowing it to considering values other
than the key field of prior inserted nodes. This concept based the
design of the ``Hidden Binary Search Tree (HBST)'', a balanced rotation-free
tree data structure. To successfully build a search path, HBST compares
the input key against ``hidden'' reference values of a reference balanced BST
ideally built on the interval $0\dots 2^B$, where $B$ is the size of nodes'
keys in bits and $2^B$ is the size of input, i.e. size $n$ of insertion keys.

We presented search, insertion and deletion algorithms that keep the HBST's height 
bounded to $O(B)$. Since $B$ is dimensioned according to the insertion sequence 
size $n$ in such a way that $B=O(\log_2 n)$, HBST achieves logarithmic worst-case
running time under the assumption that $B$ is fixed once $n$ is given. This is the 
same assumption under which AVL and Red-Black BSTs achieve $O(\log_2 n)$ worst-case time. 
In fact, as pointed out by~\cite{cormen-2009}, if $B$ can grow arbitrarily, a `simple' $O(1)$ 
key comparison (as assumed by AVL and Red-Black) becomes unrealistic, preventing those trees' 
complexities to be solely explained by $O(\log_2n)$. Under this same assumption, 
 HBST achieves $O(B)=O(\log_2 n)$ time with no need to perform any kind of balancing/rotation 
(as required by AVL and Red-Black) nor to assume special order on the input, as required by BST to
achieve logarithmic performance.

An important question left open in this work is about the feasibility of a 
linear time \emph{in-order} traversal on HBST. Regarding the presented functions, lots of 
interesting refinements can be performed such as adaptive $B$ according to the given 
key value, top-down insertion and hybrid BST-HBST structures. Finally, it would be
interesting to check whether the ``hidden search'' property can improve the performance
of other kind of structures such as external data structures (e.g. B-tree) and
priority queues.

\section{Acknowledgements}
I would like to thank Edimar Bauer, our teaching advisor for the courses of algorithms and
data structures during 2017/2. Thank you for embracing the idea in a so enthusiastic way, 
pointing out improvements and performing lots of tests!
\bibliography{refs}
\bibliographystyle{plain}
\end{document}